\documentclass[10pt,a4paper,twocolumn,italian,english,superscriptaddress,prl,nofootinbib]{revtex4}
\usepackage{lmodern}

\usepackage[T1]{fontenc}
\usepackage[latin9]{inputenc}
\usepackage{babel}
\usepackage{amsmath}
\usepackage{amssymb}
\usepackage{graphicx}
\usepackage{esint}
\usepackage[unicode=true,pdfusetitle,
 bookmarks=true,bookmarksnumbered=false,bookmarksopen=false,
 breaklinks=false,pdfborder={0 0 1},backref=false,colorlinks=false]
 {hyperref}

\makeatletter

\pdfpageheight\paperheight
\pdfpagewidth\paperwidth

\providecommand{\tabularnewline}{\\}

\@ifundefined{textcolor}{}
{%
 \definecolor{BLACK}{gray}{0}
 \definecolor{WHITE}{gray}{1}
 \definecolor{RED}{rgb}{1,0,0}
 \definecolor{GREEN}{rgb}{0,1,0}
 \definecolor{BLUE}{rgb}{0,0,1}
 \definecolor{CYAN}{cmyk}{1,0,0,0}
 \definecolor{MAGENTA}{cmyk}{0,1,0,0}
 \definecolor{YELLOW}{cmyk}{0,0,1,0}
}

\renewcommand{\[}{\begin{equation}}
\renewcommand{\]}{\end{equation}} 

\makeatother

\begin{document}

\title{Firm size distribution in Italy and employment protection}

\author{Luca Amendola}

\affiliation{Institute of Theoretical Physics, Ruprecht-Karls-Universität Heidelberg,
Philosophenweg 16, 69120 Heidelberg, Germany}
\begin{abstract}
The number of Italian firms in function of the number of workers is
well approximated by an inverse power law up to 15 workers but shows
a clear downward deflection beyond this point, both when using old
pre-1999 data and when using recent (2014) data. This phenomenon could
be associated with employent protection legislation which applies
to companies with more than 15 workers (the \emph{Statuto dei Lavoratori}).
The deflection disappears for agriculture firms, for which the protection
legislation applies already above 5 workers. In this note it is estimated
that a correction of this deflection could bring an increase from
3.9 to 5.8\% in new jobs in firms with a workforce between 5 to 25
workers.
\end{abstract}
\maketitle

\section{Introduction}

The average number of workers per company in Italy is much lower than
the EU average: 3.9 versus 6.1 \cite{istat}. This relative smallness
of Italian firms is often seen as one of the causes of the weakness
of the Italian economy in the global markets. One hypothesis that
has been advanced to explain the phenomenon is that the Italian legislation
makes it costly for a company to grow to over 15 workers because above
this threshold the legislation known as the \emph{Statuto dei Lavoratori}
(Worker's Statute) applies. This includes, in particular, the restrictions
on firing workers with open-ended contracts (the so-called \emph{Art.
18}). In the case of agriculture firms, the \emph{Statuto} applies
already above 5 workers. The strong financial compensation to the
fired employee enforced in some cases by the \emph{Art. 18}, in addition
to the risk of a forced reinstatement and a long legal litigation,
could lead some companies to choose to remain undersized \cite{key-4}.
A concise and clear history of the employment protection in Italy
can be found in \cite{italy2} and will not be repeated here.

In order to test the hypothesis that the \emph{Statuto} limits the
growth, we analyse here three sets of data, one obtained by an average
over the years 1986-1999 for all the Italian private firms between
5 and 25 workers (we denote it as pre-1999 data, \cite{key-2}) and
two for year 2014 \cite{infocamere} for the same workforce range.
The 2014 data has been collected separately for agriculture firms
(we denote this dataset as \emph{Agri2014} data) and for the rest
(we denote this as 2014 data) in order to further test the effect
of the threshold at 15 workers. Some global figure about the three
datasets are reported in Tab. \ref{tab:Firm-data}. In all cases the
workforce includes open-ended as well as temporary workers at the
time of the recording. The definition of agriculture firms in the
\emph{Agri2014} data is not completely equivalent to the legal definition
for as concerns the \emph{Statuto }so the \emph{Agri2014} data for
firms above 15 workers might be contaminated by an unknown number
of firms that are actually subject to the \emph{Statuto} provisions;\emph{
}what is important for the present analysis, however, is that all
the firms with more than 15 workers in the 2014 dataset are subject
to the \emph{Statuto}. For the pre1999 data, the agriculture firms
are included in the dataset. As one can see from the relative numbers
in Tab. \ref{tab:Firm-data}, however, the fraction of the agriculture
firm should be rather small.

\begin{figure}
\includegraphics[width=8cm]{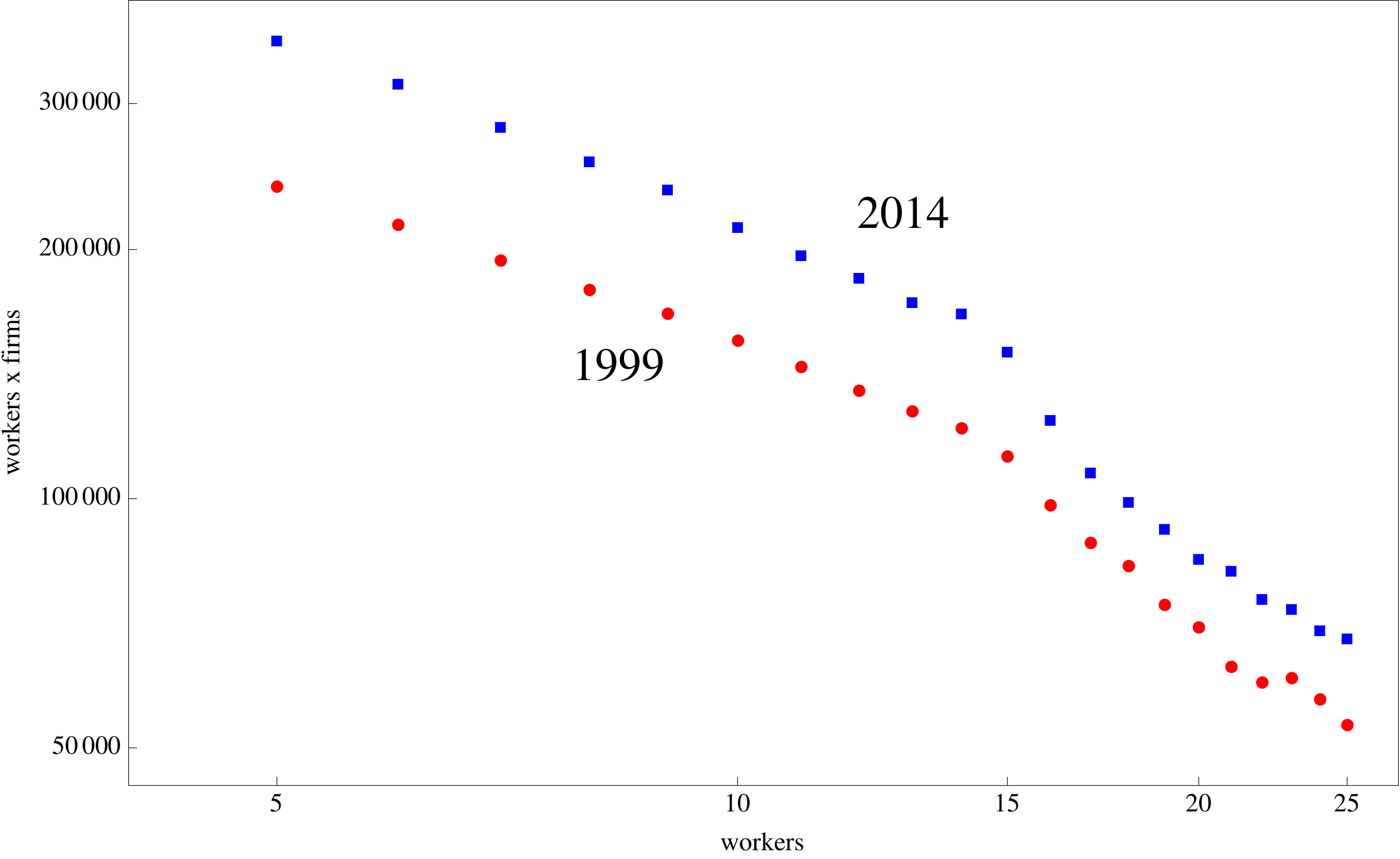}

\protect\caption{\label{fig:Number-of-workers}Total number of workers employed in
firms with a given number of workers for the pre-1999 data (red circles)
and for the 2014 data (blue squares).}
\end{figure}

\begin{figure}
\includegraphics[width=8cm]{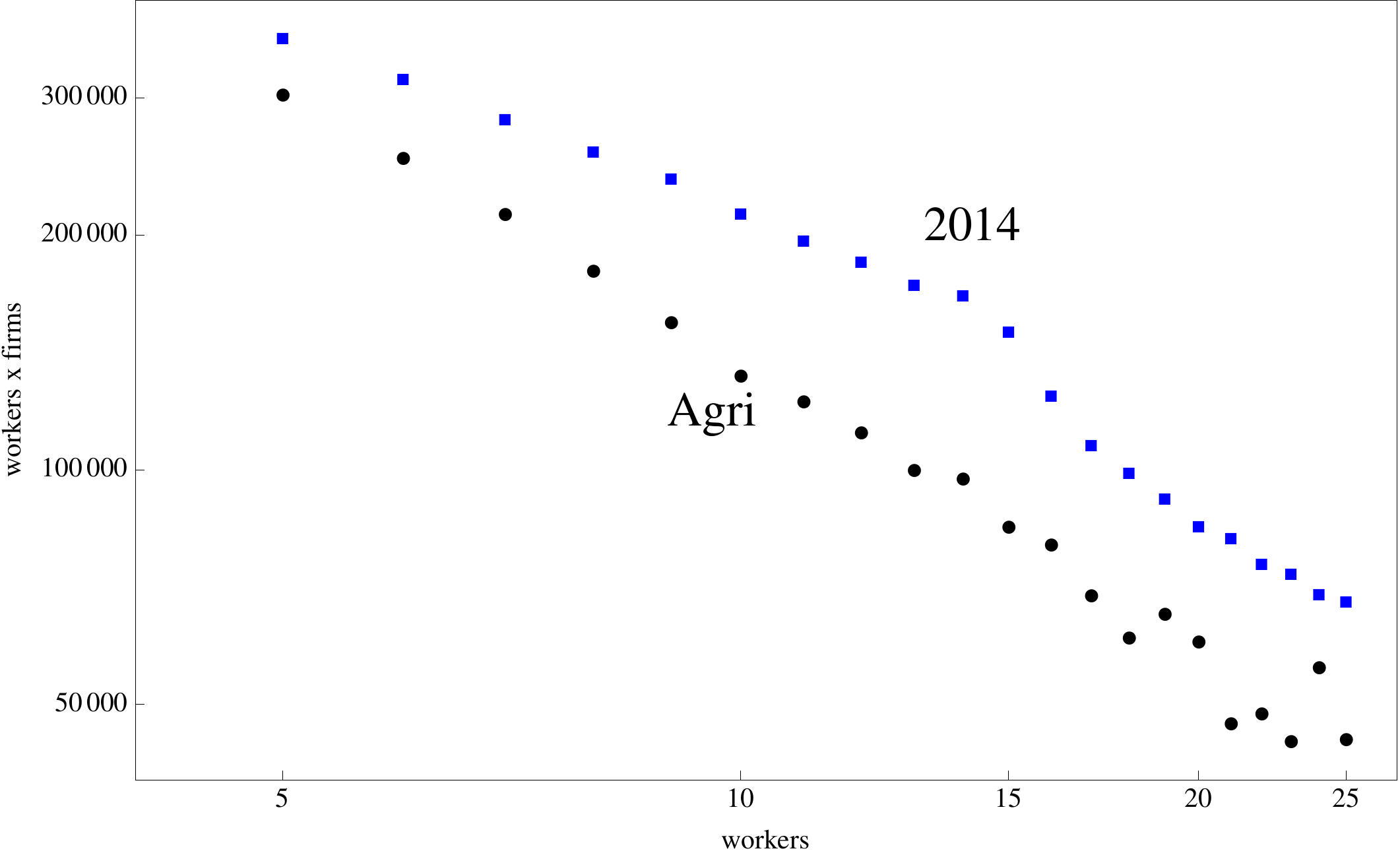}

\protect\caption{\label{fig:Number-of-workers-1}Total number of workers employed in
firms with a given number of workers for the 2014 data (blue squares)
and for the \emph{Agri2014}data (black circles).}
\end{figure}

The pre1999 and the 2014 datasets are shown in Fig. (\ref{fig:Number-of-workers})
in logarithmic coordinates. Denoting with $n(A)$ the number of firms
with $A$ workers, we show on the ordinate the total number $A\cdot n(A)$
of workers in firms with $A$ workers. The trend appear quite similar,
a part for the absolute values. In both cases, an initial straight
line (i.e. an inverse power law) gives way near 15 workers to a quite
steeper downward curve, which is also as a first approximation an
inverse power law. In Fig. (\ref{fig:Number-of-workers-1}) we compare
instead the 2014 data with the \emph{Agri2014} data, rescaled by an
arbitrary factor 12 in order to bring them closer to the 2014 data.
As it appears quite clearly, the \emph{Agri2014} data do not show
any obvious departure from a straight line, although they have a much
larger scatter due to their smaller absolute values. The visual inspection
of Figs. (\ref{fig:Number-of-workers}) and (\ref{fig:Number-of-workers-1})
seems therefore to support our initial hypothesis. The rest of this
paper is devoted to test it quantitatively and to extract some possible
estimate of how many jobs the removal or attenuation of the provisions
of the \emph{Statuto} could induce.

\begin{table}
\begin{centering}
\begin{tabular}{|c|c|c|c|}
\hline 
data & firms & approx. workers & \tabularnewline
\hline 
\hline 
pre-1999 & 253562 & 2,491,000 &  \cite{key-2}\tabularnewline
\hline 
2014 & 356602 & 3,401,000 & \cite{infocamere}\tabularnewline
\hline 
Agri & 21258 & 193,000 & \cite{infocamere}\tabularnewline
\hline 
\end{tabular}
\par\end{centering}

\protect\caption{\label{tab:Firm-data}Firm data}
\end{table}

\section{The pre-1999 data}

\begin{figure}
\includegraphics[width=8cm]{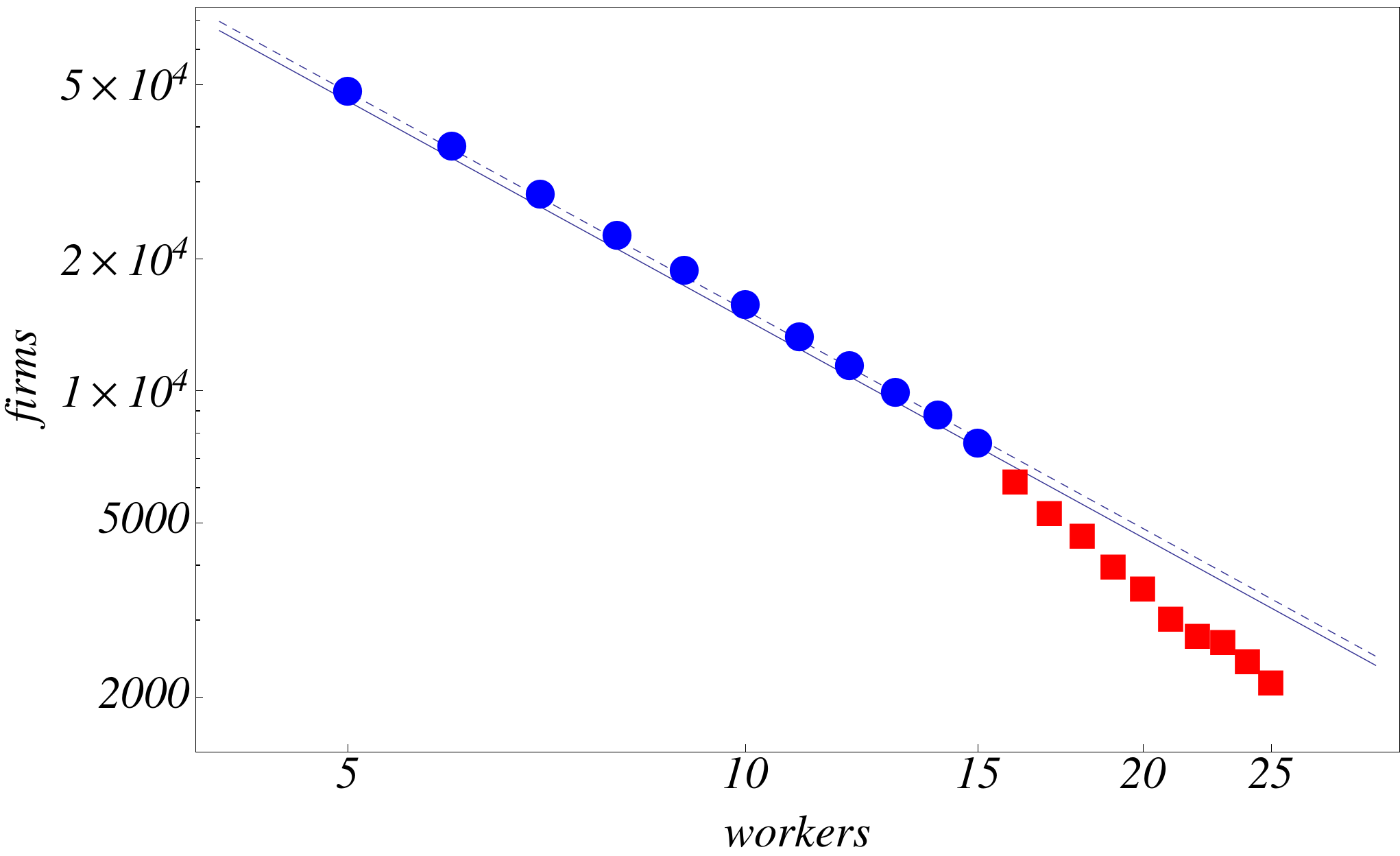}

\protect\caption{\label{fig:Numero-di-imprese}The dots represent the number of firms
as a function of the number of workers (elaboration of data from Ref.
\cite{key-2}). The dashed line indicates the best fit from 5 to 14
workers (blue circles). Subsequent points (red squares) differ significantly
from the extrapolated behavior. The continuous line is obtained by
rescaling the best fit in order to provide the same total number of
companies (fixed-slope scenario).}
\end{figure}

In a work of Schivardi and Torrini in 2008 \cite{key-2}, the pre-1999
data for the roughly 250,000 Italian companies active in the period
1986 to 1998 has been analyzed. The average $n(A)$ of the number
of firms as a function of the number of workers is shown in Fig. (\ref{fig:Numero-di-imprese})
in log-log coordinates. The authors of Ref. \cite{key-2} find that
the effect of the downward turn around 15 workers is statistically
significant but is relatively small and its removal would bring only
a 0.5\% to 1\% increase in jobs. This result was obtained by estimating
the probability of firms to hire (probability that a company increases
the number of workers) averaged over the years. This probability shows
a small dip, or anomaly, around 15 workers, that amounts to a 2\%
decrease in the probability of hiring. The hypothesis of the work
\cite{key-2} is that the effect of that depression can be corrected
by estimating what the deviation is with respect to the average performance
before and after the threshold. This method implicitly assumes that
the behavior of firms before and after the threshold is similar, that
is, not influenced by the \emph{Statuto dei Lavoratori} or other external
factors, and that an anomaly exists only in the immediate vicinity
of the transition from below to above threshold. This assumption,
however, seems at odds with Figs. \ref{fig:Number-of-workers} or
\ref{fig:Numero-di-imprese}, in which it is clear that the data do
not show a simple subsidence around $15$ workers but rather a systematic
decrease with a different trend whose effects extend to at least 25
workers. Moreover, the restrictions imposed by the \emph{Statuto }are
not fixed-cost ones since the cost of firing is proportional to the
number of workers that could potentially be fired and therefore to
the total number of workers. One should then expect an effect that
persists also for firms well above the threshold. In Ref. \cite{italy2}
the authors find indeed that the behavior of firms with respect to
the employment of temporary workers (not protected by law) changes
quite significantly below and above threshold and that this change
is detectable up to at least firms with 25 workers.

A statistically significant reduction between 1\% and 5\% growth rate
for companies with 15 workers in a sample of 2572 companies was also
obtained in \cite{estratt} but the employment consequences have not
been quantified.

As already mentioned, the pre-1999 data includes all types of private
activity, including agriculture firms to which the \emph{Statuto dei
Lavoratori} legislation applies already above 5 workers. However these
firms are typically very few and very small: the fraction of agriculture
firms with 15-25 workers is less than 1\% of the total number of firms
in our 2014 data. Therefore we can safely neglect the effect of this
component in our results.

The inflection point in Fig. (\ref{fig:Numero-di-imprese}) can be
estimated from the data itself, without prior knowledge of a certain
threshold. To this aim, we employed the RANSAC algorithm \cite{ransac},
which consists in a search of a best fit composed of two independent
straight lines on the log-log plane $\log n,\log A$. In all the simulations
performed with various probability threshold (points farther than
2 or 3$\sigma$ are rejected from the consensus fit), the intersection
between the two lines has always been correctly identified in the
vicinity of 15 workers and most of the points upstream and downstream
of the threshold have been assigned to the two different power laws.
The existence of a statistically significant change around 15 workers
is thus confirmed independently of the hypothesis that it is caused
by legislation. This implies that the firm distribution shows a discontinuity
in its first derivative rather than in the distribution itself (the
continuity of the distribution has been ascertained also in \cite{italy2}).
Given the existence of a statistically significant break, one should
fit separately the trend below and above threshold.

The curve for $A=5-15$ appears to be very well described by a power
law (Pareto distribution) obtained by the method of least squares

\begin{equation}
n_{fit}(A)=10^{5.838\pm0.006}A^{-(1.645\pm0.007)}\label{fit}
\end{equation}

This estimate includes Poisson errors in the number of firms, $\Delta n/n=1/\sqrt{n}$,
although the effect of the errors is negligible (they would appear
smaller than the symbols used in Fig.\ref{fig:Numero-di-imprese}.)
The data for $A>16$ lie systematically below the power law. The slope
of the best fit is $-1.645$ for $A=5-15,$ steepens to $-2.34$ if
one fits in the range $A=16-25$ and is $-1.82$ if one fits the entire
sequence; a power law fit applied to the entire sequence does not
appear, however, statistically acceptable. 

A similar effect was found in France \cite{france} near the threshold
of 50 workers, beyond which a number of protective regulations and
tax compliance come into effect. According to \cite{france}, the
power law $n(A)$ has a slope of -1.82 (between 10 and 1000 workers),
virtually identical to what we find in Italy between 5 and 25 workers.
In France there are however other thresholds at 10 and 20 workers
so the comparison is not straightforward \cite{france2}.

\begin{figure}
\includegraphics[width=8cm]{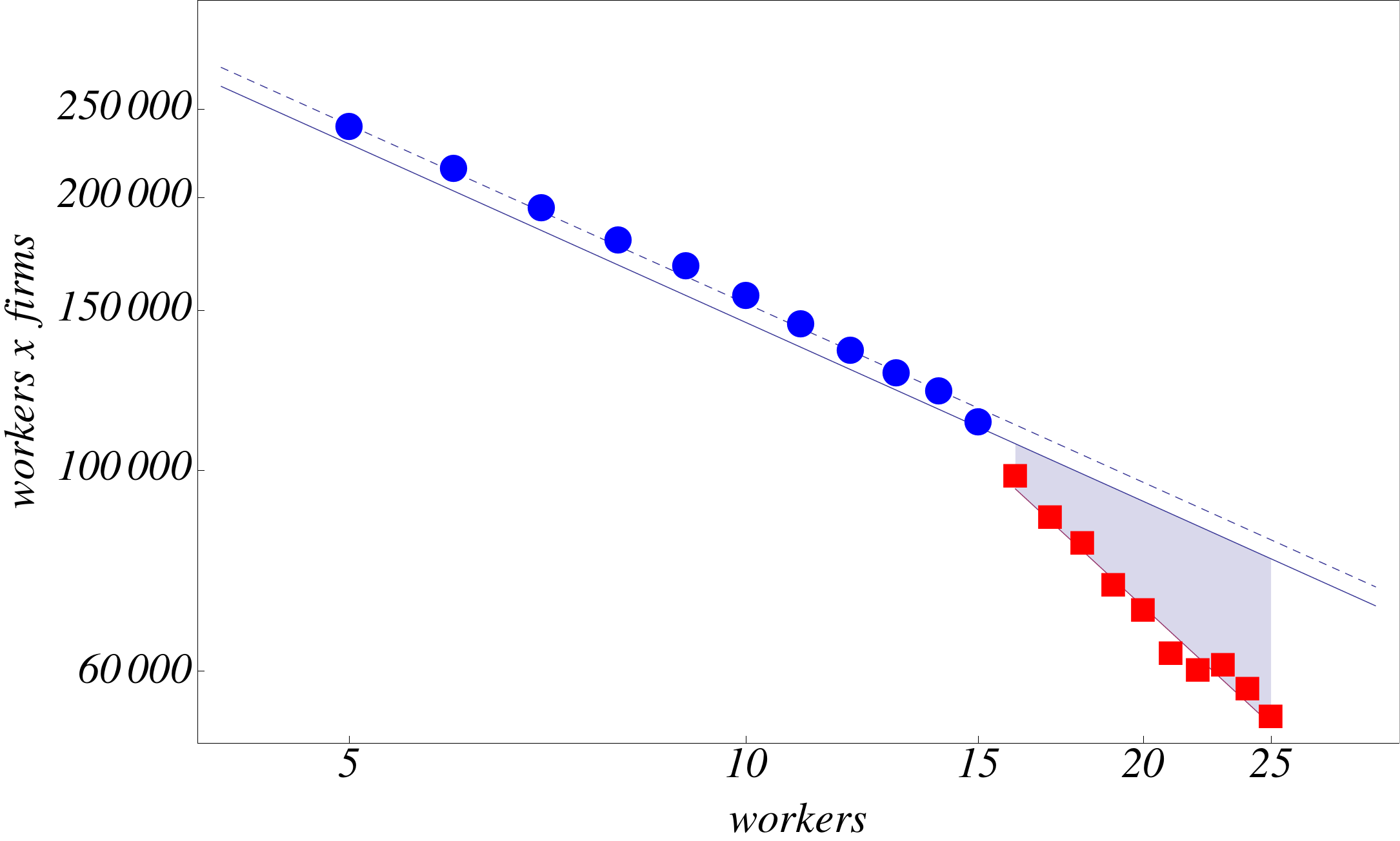}

\protect\caption{\label{fig:Totale-numero-occupati}Total number of workers per worker
bin. As in Fig. 1, the dashed line indicates the best fit from 5 to
15 workers (blue circles). The continuous line is obtained by rescaling
the best fit in order to provide the same total number of companies
(fixed-slope scenario). The grey region denotes the total number of
new workers in firms above 15 workers one would obtain if the change
in slope could be removed.}
\end{figure}

We can now try to estimate the effect of removing the causes of the
change in slope, whether they are associated to legislation or not.
To do so we make a first crucial hypothesis (\emph{power-law hypothesis}):
that the distribution of firm sizes when no artificial restriction
on growth are imposed is a power law. This hypothesis is based on
the fact that indeed a power law has been found to be a good approximation
to firm distribution in the US \cite{usa}, in France \cite{france}
(with a break around 50 as mentioned), in our own data separately
below and after the break at $A=15$. Even where a small deviation
from a power law over large size intervals has been observed, for
instance in Dutch manifacturing firms \cite{dutch}, a power law approximation
looks very reasonable in the range below 25 workers.

Once the shape of the trend is fixed, we are left with just two parameters,
normalization and slope. Since we are trying to forecast a redistribution
of firms regardless of their birth or mortality rates, we will assume
that the number of firms remain constant after the barrier removal.
This leaves a single parameter to fit, either the slope or the normalization
or a combination thereof. Next, we make a second crucial assumption,
to be referred to as the \emph{growth hypothesis}, namely, that the
number of small firms will not increase in number as a consequence
of the barrier removal. This seems very reasonable, since it would
be rather hard to explain why the number of small firms should increase
when a barrier to \emph{growth} has been removed. One could argue
perhaps that the current legislation is also a barrier to decrease,
because it forces the firms to retain workers that would otherwise
be fired. This explanation is however untenable, first because a firm
could anyway reduce its employment by simply ceasing the natural turn-over
and secondly because a firm will anyway fire its workers if the financial
conditions require it, or else it would fail. If we assume then that
the small firms do not increase in number, then there are only two
possible limiting scenarios, that now we discuss in turn.

In the first scenario (let us call it \emph{fixed-slope scenario}),
we can assume that the firms with less than 15 workers have been currently
growing unimpeded as long as they stayed below 15 workers. If this
were the case then, by removing the barrier, the behavior observed
below 15 workers would extend to above the threshold. If in addition
as already emphasized we keep constant the number of firms, then the
after-removal behavior would be represented by an extrapolation of
the same $-1.645$ power law to $A=25$ accompanied by a rescaling
to match the total number of firms. Then we could estimate an additional
number of workers among firms with $A>15$ equal to the area between
the best fit \ref{fit} and the observed data of Fig. \ref{fig:Totale-numero-occupati},
accompanied by a decrease of firms and therefore of workers employed
in firms below threshold. If we denote by $n(A_{i})$ the number of
firms with $A_{i}$ workers, we obtain an estimate of an additional
number of workers equal to 
\begin{equation}
\Delta A_{tot}=\sum_{i=5}^{25}A_{i}[\alpha n_{fit}(A_{i})-n(A_{i})]\approx130,000\pm5000
\end{equation}
where $\alpha\approx0.95$ takes into account the rescaling needed
so that the total number of companies is kept equal to the original
one. The number of additional workers represents about 5\% of the
total two and a half million workers of the pre-1999 sample. The number
of companies with 25 workers would increase from about 2100 to about
3200. We are of course making the tacit assumption that there is no
significant shortage of workers to hire. The estimate slightly increases
if to stay further away from the theshold we take as natural behavior
the trend up to 14 workers ($136,000$ new jobs) or 13 ($139,000$
new jobs). These results and the others to be discussed below are
summarized in Tab. (\ref{tab:Summary-of-results}).

In the second, more conservative, scenario (the \emph{fixed-normalization
scenario}), we can assume that the firms with the lowest number of
workers would be unaffected by the barrier removal and would stay
the same. Then in order to match the total number of firms we should
change the slope of the power law. This is obtained by a slope approximately
equal to $-1.72$, i.e. slightly steeper than before. Using this power
law as the after-removal distribution, we estimate a number of new
workers around $89,000\pm5000$, i.e. 3.6\% of the total sample. Notice
that this result is not an extrapolation of the below-threshold behavior:
it is entirely based on the power-law and the growth hypotheses.

Any other scenario would have either to lie above the fixed-normalization
scenarios or to violate one of the two basic assumptions: either it
should abandon the hypothesis of a power law, or lead to a number
of small firms that increases when the barrier is removed. In other
words, given the assumptions, the fixed-normalization scenario produces
the most conservative estimate of the number of additional workers
while we can take the fixed-slope case as a reasonable upper limit.

\section{The 2014 data}

\begin{figure}
\includegraphics[width=8cm]{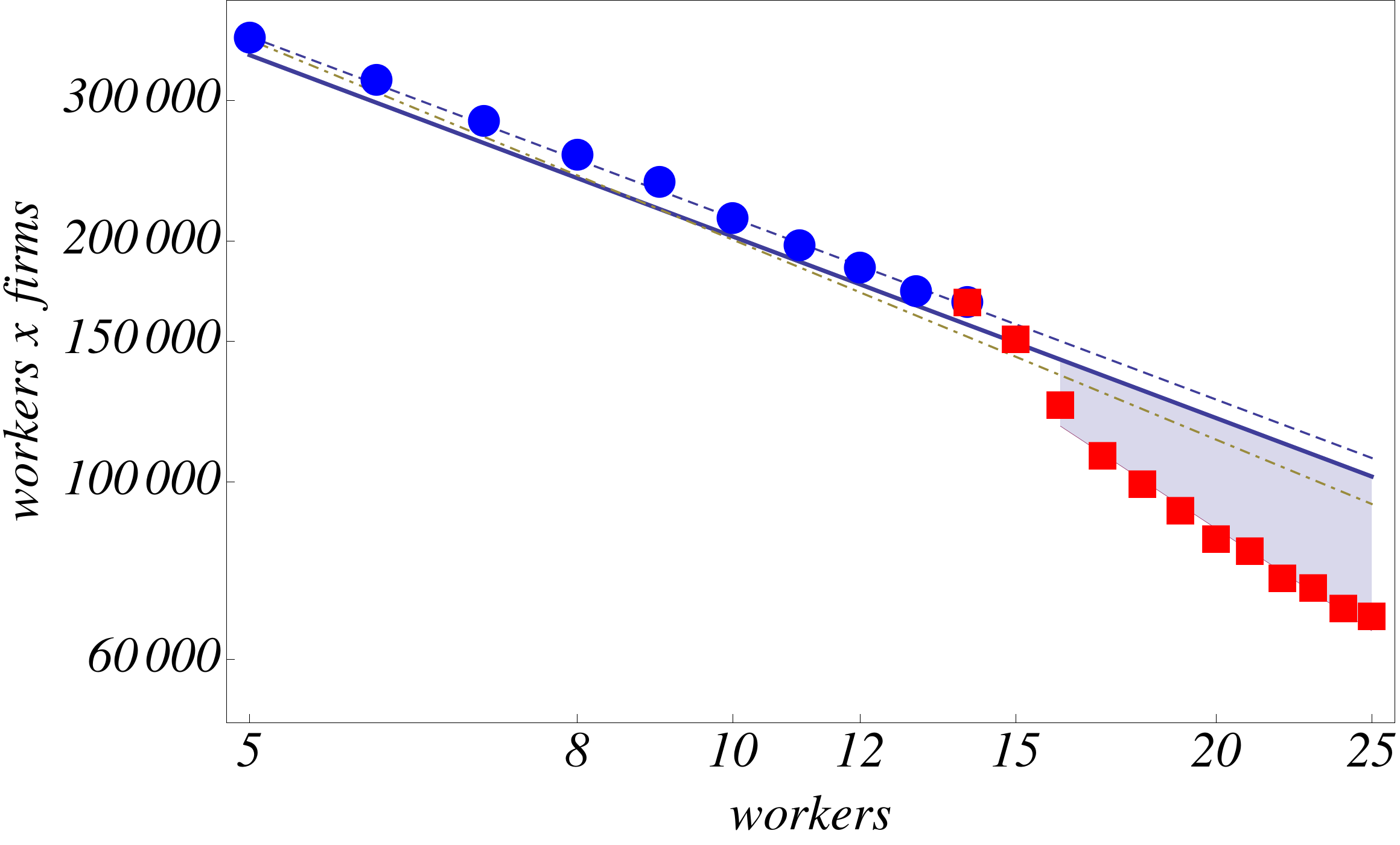}

\protect\caption{\label{fig:Totale-numero-occupati-1}Same as Fig. \ref{fig:Totale-numero-occupati},
now for the 2014 data. Here the dot-dashed line represents the fixed-normalization
scenario.}
\end{figure}

We repeated the analysis using the 2014 data, consisting of roughly
357,000 private non-agriculture firms between 5 and 25 workers \cite{infocamere}.
Here again the distribution is characterized by two different slopes
(see Fig. \ref{fig:Totale-numero-occupati-1}), slightly steeper than
for the pre-1999 data: $-(1.75\pm0.006)$ for $A=5-14$ and $-2.32$
for $A=16-25$ (or $-2.21$ for $A=17-20$). The broken power-law
distribution remains qualitatively the same even when decomposing
the data along regional units. Using the fixed-slope scenario we estimate
198,000 new jobs, i.e. a 5.8\% increase while we obtain 132,000 new
jobs (3.9\%) with the fixed-normalization scenario. The fixed-slope
figure remains practically the same (197,000 jobs) if we employ the
fit up to 13 workers. In both cases the error is estimated around
5000 units. See the summary in Tab. (\ref{tab:Summary-of-results})
for additional cases.

\begin{figure}
\includegraphics[width=8cm]{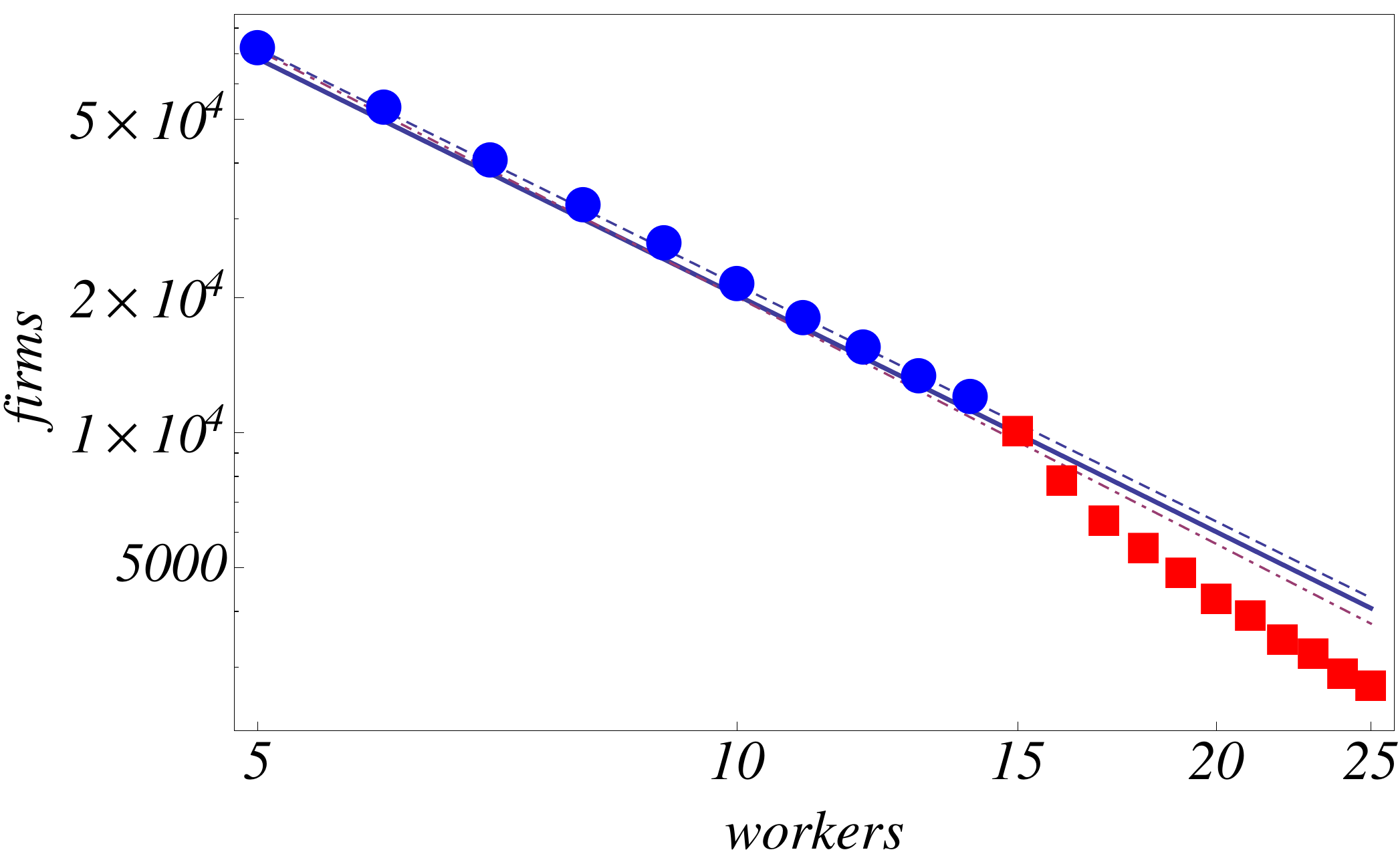}

\protect\caption{\label{fig:Totale-numero-occupati-1-1}Same as Fig. \ref{fig:Totale-numero-occupati},
now for the 2014 data. Here the dot-dashed line represents the fixed-normalization
scenario.}
\end{figure}

\section{The agriculture firms }

Contrary to the other cases, the \emph{Agri2014} dataset can be fitted
by a single power law with slope $-(2.19\pm0.025)$: the $\chi^{2}$
is 26.9 for 19 degrees of freedom ($p$-value 0.1). Adopting the fixed-slope
scenario and extrapolating the slope between $5-14$ workers, we obtain
an estimated increase of $1500\pm1100$ jobs, i.e. $0.8\pm0.6$\%,
which reduces to $0.5\pm0.5$\% with the fixed-normalization. This
confirms the visual impression from Fig. \ref{fig:Number-of-workers-1}
that the agriculture firms do not show a significant change around
15 workers.

\begin{table}
\begin{centering}
\begin{tabular}{|c|c|c|c|c|c|}
\hline 
data & polyn. order & fit range & method & jobs & \%\tabularnewline
\hline 
\hline 
pre-1999 & 1 & 5-15 & FS & 130,000 & 5.2\tabularnewline
\hline 
pre-1999 & 1 & 5-14 & FS & 136,000 & 5.4\tabularnewline
\hline 
pre-1999 & 1 & 5-13 & FS & 139,000 & 5.6\tabularnewline
\hline 
pre-1999 & 1 & - & FN & 89,000 & 3.6\tabularnewline
\hline 
pre-1999 & 2 & 5-14 & FS & 103,000 & 4.1\tabularnewline
\hline 
pre-1999 & 3 & 5-14 & FS & 88,000 & 3.5\tabularnewline
\hline 
pre-1999 & 4 & 5-14 & FS & 120,000 & 4.8\tabularnewline
\hline 
2014 & 1 & 5-14 & FS & 198,000 & 5.8\tabularnewline
\hline 
2014 & 1 & 5-13 & FS & 197,000 & 5.8\tabularnewline
\hline 
2014 & 1 & - & FN & 132,000 & 3.9\tabularnewline
\hline 
2014 & 2 & 5-14 & FS & 163,000 & 4.8\tabularnewline
\hline 
2014 & 3 & 5-14 & FS & 226,000 & 6.6\tabularnewline
\hline 
2014 & 4 & 5-14 & FS & 446,000 & 13.0\tabularnewline
\hline 
Agri2014 & 1 & 5-14 & FS & 1500 & 0.8\tabularnewline
\hline 
Agri2014 & 1 & - & FN & 900 & 0.5\tabularnewline
\hline 
\end{tabular}
\par\end{centering}

\protect\caption{\label{tab:Summary-of-results}Summary of results (FS=fixed normalization;
FS= fixed slope). The relative error on the number of jobs is around
5\% except for the last two rows, where it is around 70-100\%.}

\end{table}

\section{Further tests}

The analysis can be repeated under slightly different assumptions. 

The error attributed to the Poissonian counts is probably too optimistic,
since it provides a $\chi^{2}$ statistic per degree of freedom for
the pre-1999 data in the range $A=5-15$ equal to 3.8 ($p$-value
0.0001). Increasing the errors by a factor of 1.9 produces a $\chi^{2}$
per degree of freedom equal to about 1. The fit $n_{fit}(A)$ remains
the same but the uncertainty on $\Delta A_{tot}$ increases slightly
from 5000 to 8000. Removing the entry at $A=15$ from the fit, the
$p$-value increases to 0.02, which could indicate that the $A=15$
entry is slightly anomalous. This is hardly surprising, since it is
an average obtained over several years and therefore will include
firms that for some time have been above threshold; moreover, the
very definition of ``number of workers'' has been fluctuating over
time and a firm might well decide to stay well below the threshold
to avoid legal complications. The same factor 1.9 also brings the
linear fit for the 2014 data to a $\chi^{2}$ per degree of freedom
close to unity. Note that since the workforce includes temporary workers
at the time of recording, the average number of workers in a firm
fluctuates around the given values.

The curve data $\log n_{i}$ $vs$ $\log A_{i}$ in the range $A=5-14$
can be approximated also by a polynomial of higher order rather than
by a straight line. With a polynomial of order 2,3 or 4 we obtain
respectively $\Delta A_{tot}=$ 103000, 88000, 120000 for the pre-1999
data. Note however that in these cases, assuming Poisson errors, the
$\chi^{2}$ per degree of freedom becomes only 0.5-0.7 and obviously
decreases with more realistic (i.e. larger) errors. Higher order polynomials
therefore do not seem to be supported by the data in the range $A=5-14$.
Higher order polynomial are instead statistically acceptable fits
to the 2014 data, but then $\Delta A_{tot}$ is close to or larger
than  the straight line result (see Tab. \ref{tab:Summary-of-results}).

\section{Conclusions}

From this brief analysis we cannot of course derive in a mechanical
way a causal relationship between the broken power law $n(A)$ and
the \emph{Statuto dei Lavoratori}. It is also impossible from this
data to establish what other side effects a reform of the \emph{Statuto}
could have, for instance whether it will induce merging of small companies
rather than real growth or whether it will modify the employment of
temporary versus open-ended workers.

It seems clear, however, that the behavior of firms does change in
correspondence of 15 workers. Removing the causes of such behavior,
whatever they may be, could generate from roughly 130,000 to 200,000
new workers in our sample, about 3.9-5.8\% of the total 2014 sample.
These figures are, it should be noted, only a lower estimate, because
limited due to lack of data to companies with fewer than 25 workers.
\begin{acknowledgments}
\emph{Acknowledgments}. I am thankful to Roberto Susanna (Ufficio
stampa InfoCamere infocamere.it) for providing the 2014 dataset. I
acknowledge very useful comments from Melanie Arntz (University of
Heidelberg), Andrea Bassanini (OECD), Carlo Clericetti, Luca Ferretti,
Carlo Menon (OECD), Fabiano Schivardi (LUISS University), Giuseppe
Vandai, Katharina Wetzel-Vandai.\end{acknowledgments}


\begin{thebibliography}{10}
\bibitem{infocamere}Data obtained from the Italian firm registry
infocamere.it.
\end{thebibliography}
\end{document}